\documentclass[twocolumn]{article}
\usepackage{cite}
\usepackage{graphicx}
\usepackage{amsmath}
\usepackage{authblk}
\usepackage[english]{babel}
\usepackage{blindtext}
\usepackage[utf8]{inputenc}
\usepackage[binary-units=true]{siunitx}
\usepackage{dirtytalk}
\usepackage{url}

\title{\bf{Machine Learning Pipelines with Modern Big Data Tools for High Energy Physics}}
\author[1,2]{M. Migliorini}
\author[1]{R. Castellotti}
\author[1,*]{L. Canali}
\author[2]{M. Zanetti}
\affil[1]{European Organization for Nuclear Research (CERN), Geneva, Switzerland}
\affil[2]{University of Padova, Padua, Italy}

\affil[*]{luca.canali@cern.ch}

\begin{document}
\date{}
\maketitle 

\begin{abstract}

The effective utilization at scale of complex machine learning (ML) techniques for HEP use cases poses several technological challenges, most importantly on the actual implementation of dedicated end-to-end data pipelines. A solution to these challenges is presented, which allows training neural network classifiers using solutions from the Big Data and data science ecosystems, integrated with tools, software, and platforms common in the HEP environment. In particular, Apache Spark is exploited for data preparation and feature engineering, running the corresponding (Python) code interactively on Jupyter notebooks. Key integrations and libraries that make Spark capable of ingesting data stored using ROOT format and accessed via the XRootD protocol, are described and discussed. Training of the neural network models, defined using the Keras API, is performed in a distributed fashion on Spark clusters by using BigDL with Analytics Zoo and also by using TensorFlow, notably for distributed training on CPU and GPU resourcess. The implementation and the results of the distributed training are described in detail in this work.

\end{abstract}

\section*{Introduction}

High energy physics (HEP) experiments like those at the Large Hadron Collider (LHC) are paramount examples of ``big-data'' endeavors: chasing extremely rare physics processes requires producing, managing and analyzing large amounts of complex data. Data processing throughput of those experiments is expected to exceed 200 TB/s in 2025 with the upgrade of the LHC (HL-LHC project), which, after tight online filtering, implies saving on permanent storage 100 PB of data per year. Thorough and complex data processing enables major scientific achievements, like the discovery of the Higgs boson in 2012. Data ingestion, feature engineering, data reduction and classification are complex tasks, each requiring advanced techniques to be accomplished. 
While, so far, custom solutions have been employed, recent developments of open source tools are making the latter compelling options for HEP specific use-cases. Furthermore, physics data analysis is profiting to a large extent from modern Machine Learning (ML) techniques, which are revolutionizing each processing step, from physics objects reconstruction (feature engineering), to parameter estimation (regression) and signal selection (classification).
In this scope, \textit{Apache Spark}\cite{Zaharia:2010:SCC:1863103.1863113} represents a very promising tool to extend the traditional HEP approach, by combining in a unique system powerful means for both sophisticated data engineering and Machine Learning.  
Among the most popular analytics engines for big data processing, Spark 
allows performing interactive analysis and data exploration through its mature data processing engine and API for distributed data processing, its integration with cluster systems and by featuring ML libraries giving the possibility to train in a distributed fashion all common classifiers and regressors on large datasets.

It has been proved (e.g. in \cite{Baldi:2014kfa}) that Deep Learning can boost considerably the performances of physics data analysis, yielding remarkable results from larger sets of low-level (i.e. less ``engineered'') quantities. There is thus a significant interest in integrating Spark with tools, like BigDL\cite{2018arXiv180405839D}, allowing distributed training of deep learning models.

The development of an end-to-end machine learning pipeline to analyze HEP data using Apache Spark is described in this paper. After briefly recalling the traditional data processing and analysis workflow in HEP, the specific physics use-case addressed in work is presented; the various steps of the pipeline are then described in detail, from data ingestion to model training, whereas the overall results are reported in the final section. 

The primary goal of this work is to reproduce the classification performance results of 
\cite{2018arXiv180700083N} using tools from the Big Data and data science ecosystems, showing that the proposed pipeline makes more efficient usage of computing resources and provides a more productive interface for the physicists, along all the steps of the processing pipeline.

\section*{Traditional Analysis Workflow and Tools}

The backbone of the traditional HEP analysis workflow is ROOT\cite{BRUN199781}, a multipurpose \textit{C++} toolkit developed at CERN implementing functionalities for I/O operations, persistent storage, statistical analysis, and data visualization.
Data gathered by LHC experiments or produced by their simulation software are provided in ROOT format, with file-based data representation and an event-based class structure with branches. The latter is a feature of paramount importance, as it enables the flexibility required to preserve the complexity of the recorded data, allowing keeping track of intrinsic dependencies among physics objects of each collision event.

Centralized production systems orchestrate the data processing workflow, converting the raw information into higher-level quantities (data or feature engineering). Computing resources, organized world-wide in hierarchical tiers, are exploited employing GRID protocols \cite{Bird:1695401}. Although the centrally produced datasets may require some additional feature preparation, from this stage on, the processing is analysis-dependent and is done by the users using batch systems. Machine Learning algorithms are executed either from within the ROOT framework (using TMVA\cite{TMVA}, a toolkit for multivariate data analysis) or using more common open source frameworks (Keras/Tensorflow, PyTorch, etc.).


\section*{The Physics Use Case}

Physicists primarily aim at distinguishing interesting collision events from uninteresting ones, the former being those associated with specific physics signals whose existence is sought or whose properties are worth being studied. 
Typically, those signals are extremely rare and correspond to a tiny fraction of the whole dataset.
Data analysis results then in a classification problem, where, in addition to the signal category, the background is often also split into several classes.  

Out of the 40 million collisions produced by the LHC every second, only a small fraction (about 1000 events) can currently be stored by the online pipelines of the two omni-purpose detectors, CMS and ATLAS. A haphazard selection of those events would dilute the already rare signal processes, thus efficient data classification needs to take place already online. Experiments implement complex trigger systems, designed to maximize the true-positive rate and minimize the false-positive rate thus allowing effective utilization of computing resources both online and offline (e.g. processing units, storage, etc.).

The Machine Learning pipeline described in this paper addresses the same physics use-case considered in the work by Nguyen et al. \cite{2018arXiv180700083N} where
event topology classification, based on deep learning, is used to improve the purity of data samples selected at trigger level. The dataset is the result of a Monte Carlo event generation, where three different processes (categories) have been simulated: the inclusive production of a leptonically decaying $W^{\pm}$ boson, the pair production of a top-antitop pair ($t\bar{t}$), and hadronic production of multijet events. Variables of low and high level are included in the dataset.

\section*{Data Pipeline For Machine Learning}

\begin{figure}
\centering
\includegraphics[width=3in]{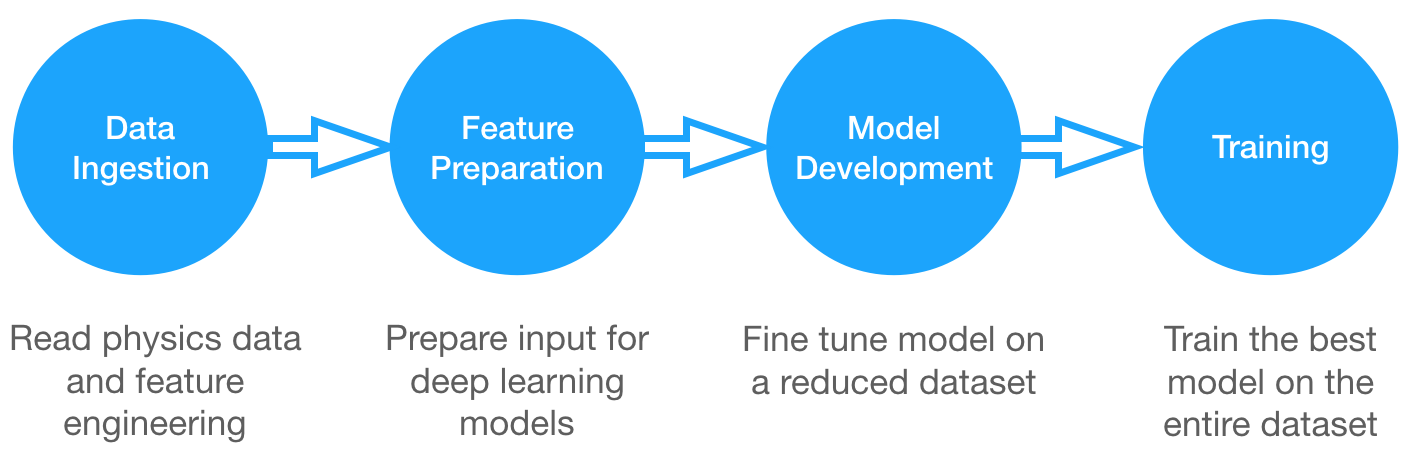}
\caption{Scheme of the data processing pipeline used for training the event topology classifier.}\label{ref:DataPipeline}
\end{figure}

Data pipelines are of paramount importance to make machine learning projects successful, by integrating multiple components and APIs used across the entire data processing chain. A good data pipeline implementation can help to achieve analysis results faster by improving productivity and by reducing the amount of work and toil around core machine learning tasks. In particular, data pipelines are expected to provide solid tools for data processing, a task that ends up being one of the most time-consuming for physicists, and data scientists in general, approaching data analysis problems.
Traditionally, HEP has developed custom tools for data processing, which have been successfully used for decades. Recently, a large range of solutions for data processing and machine learning have become available from open source communities. The maturity and adoption of such solutions continue to grow both in industry and academia. Using software from open source communities comes with several advantages, including lowering the cost of development and support, and the possibility of sharing solutions and expertise with a large user community.
In this work, we implement the machine learning pipeline detailed in \cite{2018arXiv180700083N} using tools from the ``Big Data'' ecosystem.
One of the key objectives for the machine learning data pipeline is to transform raw data into more valuable information used to train the ML/DL models. Apache Spark provides the backbone of the pipeline, from the task of fetching data from the storage system to feature processing and feeding training data into a DL engine (BigDL and Analytics Zoo are used in this work). Additionally, distributed deep learning results are detailed, obtained using TensorFlow on cloud resources and container-based environments. The four steps of the pipeline we built are (see also \figurename{ \ref{ref:DataPipeline}}):

\begin{itemize}
    \item \textbf{Data Ingestion}, where we read data in ROOT format from the CERN EOS storage system, into a Spark DataFrame and save the results as a large table stored in a set of Apache Parquet files.
    \item \textbf{Feature Engineering and Event Selection}, where the Parquet files containing all the events details processed in Data Ingestion are filtered, and datasets with new features are produced.
    \item \textbf{Parameter Tuning}, where the hyperparameters for each model architecture are optimized by performing grid search.
    \item \textbf{Training}, where the neural network models are trained on the full dataset.
\end{itemize}

\noindent In the next sections, we will describe in detail each step of this pipeline.

\subsection{Data Source}
Data used for this work have been generated using software simulators to generate events and to calculate the detector response, as previously discussed, see also \cite{2018arXiv180700083N} for details. For this exercise, the generated training data amounts to 4.5 TB, for a total of 54 million events, divided in 3 classes: ``W + jet'', ``QCD'', ``t$\bar{\text{t}}$'' events. The generated training data is stored using the ROOT format, as it is a common format for HEP. Data are originally stored in the CERN EOS storage system as it is the case for the majority of HEP data at CERN at present. The authors of \cite{2018arXiv180700083N} have kindly shared the training data for this work. Each event of the dataset consists of a list of reconstructed particles. Each particle is associated with features providing information on the particle cinematic (position and momentum) and on the type of particle.

\subsection{Data Ingestion}
Data ingestion is the first step of the pipeline, where we read ROOT files from the CERN EOS \cite{EOS} storage system into a Spark DataFrame. For this, we use a dedicated library able to ingest ROOT data into Spark DataFrames: spark-root\cite{viktor_khristenko_jim_pivarski_2017}, an Apache Spark data source for the ROOT file format. It is based on a Java implementation of the ROOT I/O libraries, which offers the ability to translate ROOT files into Spark DataFrames and RDDs. To access the files stored in the EOS storage system from Spark applications, another library was developed: the Hadoop-XRootD connector\cite{xRootDconnector}. The Hadoop-XRootD connector is a Java library extending the Apache Hadoop Filesystem \cite{HadoopYARN} to allow accessing files stored in EOS via the XRootD protocol. This allows Spark to read directly from the EOS storage system, which is convenient for our use case as it avoids the need for copying data into HDFS or other storage compatible with Spark/Hadoop libraries.
At the end of the data ingestion step, the result is that data, with the same structure as the original ROOT files, are made available as a Spark DataFrame on which we can perform event selection and feature engineering.

\subsection{Event Selection and Feature Engineering}
In this step of the pipeline, we process the dataset by applying relevant filters, by computing derived features and by applying data normalization techniques.
The first part of the processing requires domain-specific knowledge in HEP to simulate trigger selection: this is emulated by requiring all the events to include one isolated electron or muon with transverse momentum ($p_T$) above a given threshold, $p_T \geq$\SI{23}{\giga\eV}. All particles passing the trigger selection are then ranked in decreasing order of $p_T$. For each event, the isolated lepton is the first entry of the list of particles. Together with the isolated lepton, the first 450 charged particles, the first 150 photons, and the first 200 neutral hadrons have been considered, for a total of 801 particles with 19 features each. The result is that each event is associated with a matrix with shape $801 \times 19$. This defines the Low-Level Features (LLF) dataset.
Starting from the LLF, an additional set of 14 High-Level Features (HLF) is computed. These additional features are motivated by known physics and data processing steps and will be of great help for improving the neural network model in later steps.
LLF and HLF datasets, computed as described above, are saved in Apache Parquet \cite{ApacheParquet} format: the amount of training data is reduced at this point from the original 4.5 TB of ROOT files to 950 GB of snappy-compressed Parquet files.
Additional processing steps are performed and include operations frequently found when preparing training data for classifiers, notably:
\begin{itemize}
    \item Data undersampling. This is done to work around the class imbalance in the original training data. After this step, we have the same number of events (1.4 million events) for each of the three classes.
    \item Data shuffling. This is standard practice, useful to improve the convergence of mini-batch gradient descent-based training. Shuffling is done at this stage over the whole data set, making use of Spark's ability to perform large sorts.
    \item All the features present in the datasets have been pre-processed, by scaling them take values between 0 and 1 (using MinMaxScaler) or normalized, using the StandardScaler, as needed for the different classifiers.
    \item The datasets, containing HLF and LLF features and labels, have been split into training and test datasets (80\% and 20\% respectively) and saved in two separate folders, as Parquet files. Smaller datasets, samples of the full train and test datasets, have also been generated for test and development purposes.
\end{itemize}
Training and test datasets were saved as files in snappy-compressed Parquet format at the end of this stage. 
To use TensorFlow with the data prepared in our pipeline we had to introduce an additional step, where we converted the training and test data set into TFRecord format. TFRecord file format is a simple record-oriented binary format, that in essence consists of serialized protocol buffer entries\cite{protobuf}. We used Apache Spark to convert the training and test datasets from Parquet to TFRecord using the library and Spark datasource "spark-tensorflow-connector", which is distributed as open source software by the Tensorflow project. The conversion took just a few minutes, as we ran it on a cluster, taking advantage of Spark's parallel processing capabilities.
Test and training data after this stage amount to about 250 GB. The decrease in total data size from the previous step is mostly due to the undersampling step and the fact that the population of the 3 topology classes in the original training data used for this work is not balanced.

\subsection{Neural Network Models}
We have tested three different neural network models, following\cite{2018arXiv180700083N}:
\begin{itemize}
  \item The first and simplest model is the ``HLF Classifier''. It is a fully connected feed-forward deep neural network taking as input the 14 high-level features. The chosen architecture consists of three hidden layers with 50, 20, 10 nodes activated by Rectified Linear Units (ReLU). The output layer consists of 3 nodes, activated by the Softmax activation function.
  \item The ``Particle Sequence Classifier'' is trained using recursive layers, taking as input the 801 particles in the Low-Level Features dataset. The list of particles is ordered before feeding it into a recurrent neural network.
  Particles are ordered by decreasing $\Delta R$ distance from the isolated lepton, calculated as \begin{equation*}
    \Delta R = \sqrt{\Delta \eta^2 + \Delta \phi^2}
  \end{equation*}
  where $\eta$ is the pseudorapidity and $\phi$ the azimuthal angle of the particle. 
  Gated Recurrent Units (GRU) have been used to aggregate the particles input sequence, using a recurrent layer of width 50. The output of the GRU layer is fed into a fully connected layer with 3 softmax-activated nodes. Notably, this model does not make use of High-Level Features, but only uses ``Low-Level Features'' from simulation data.
  \item The ``Inclusive Classifier'' is the most complex and complete of the 3 models tested. This classifier combines the ``HLF classifier''
  with the ``Particle Sequence Classifier''. The model consists in concatenating the 14 High Level Features to the output of the GRU layer after a dropout layer. An additional dense layer of 25 nodes is introduced before the final output layer consisting of 3 nodes, activated by the softmax activation function.
\end{itemize}

\subsection{Parameter Tuning}

Hyperparameter tuning is a common step for improving machine learning pipelines. In this work, we have used Spark to speed up this step. Spark allows training multiple models with different parameters concurrently on a cluster, with the result of speeding up the hyperparameter tuning step. We used AUC, the Area Under the ROC curve, as the performance metric to compare different classifiers. When performing a grid search each run is independent of the others, hence this process can be easily parallelized and scales very well.
For example, we tested the High-Level Features classifier, a feed forward neural network, taking as input the 14 High-Level Features. For this model, we tested changing the number of layers and units per layer, the activation function, the optimizer, etc. As an experiment, we ran grid search on a small dataset containing 100K events sampled from the full dataset, using a grid of ~200 hyper-parameters sets (models).
Hyperparameter tuning can similarly be repeated for all three classifiers described above. For the following work, we have decided to use the same models like the ones presented in \cite{2018arXiv180700083N}, as they were offering the best trade-off between performances and complexity.
To run grid search in parallel, we used Spark with spark-sklearn and the TensorFlow/Keras wrapper for scikit-learn \cite{scikitlearn}.  
Additionally, we obtained similar results for parallelizing hyperparameter search using keras-tuner \cite{kerastuner}, a specialized package for hyperparameter tuning suitable for TensorFlow and we have integrated it with cloud resources on Kubernetes with a custom script. 

\subsection{Distributed Training with Spark, BigDL/Analytics Zoo and TensorFlow}
There are many suitable software and platform solutions for deep learning training nowadays, however choosing among them is not straightforward, as many products are available with different characteristics and optimizations for different areas of application. For this work, we wanted to use a solution that easily integrates with the Spark service at CERN, running on Hadoop YARN\cite{HadoopYARN} clusters, and more recently also running Spark on Kubernetes\cite{Kubernetes} using cloud resources. GPUs and other hardware accelerators are only available in limited quantities at CERN at the time of this work, so we also wanted to explore a solution that could scale on CPUs. Moreover, we wanted to use Python/PySpark and well-known APIs for neural network processing: the Keras API\cite{chollet2015keras} in this case.
Those reasons combined with an ongoing collaboration between CERN openlab\cite{openlab} and Intel has led us to test and develop this work using BigDL\cite{2018arXiv180405839D} and Analytics Zoo\cite{AnalyticsZoo} for distributed model training. BigDL and Analytics Zoo are open source projects distributed under the Apache 2.0 license. They provide a distributed deep learning framework for Big Data platforms and are implemented as libraries on top of Apache Spark. Analytics Zoo, in particular, provides a unified analytics and AI platform that seamlessly unites Spark, TensorFlow, Keras, and BigDL programs into an integrated pipeline. Notably, with Analytics Zoo and BigDL users can work with models defined with Keras and Tensorflow\cite{Tensorflow} APIs and run the training at scale using Spark. More information on Analytics Zoo can be found in the Analytics Zoo repository \cite{AnalyticsZoo}. 
BigDL provides data-parallelism to train models in a distributed fashion across a cluster using synchronous mini-batch Stochastic Gradient Descent. Data are automatically partitioned across Spark executors as an RDD of {\it Sample}: an RDD of N-Dimensional array containing the input features and label. Distributed training is implemented as an iterative process, thus there will be multiple iterations over the same data. Reading data from disk multiple times is slow, for this reason, BigDL exploits the in-memory capabilities of Spark to cache the train RDD in the memory of each worker allowing faster access during the iterations (this also means that sufficient memory needs to be allocated for training large datasets). More information on BigDL architecture can be found in the BigDL white paper\cite{2018arXiv180405839D}.
In addition, we have tested distributed training using TensorFlow. TensorFlow version 1.14 and higher introduce an easy-to-use API for running distributed training. Using the package tf.distribute, bundled with the TensorFlow distribution, distributed training of a Keras model can be activated by simply wrapping the code defining and compiling the model with the chosen strategy for distributed training. In this work, we have used the ``Multi Worker Mirrored Strategy'' to distribute training across multiple machines. Multi worker mirrored strategy is an experimental API in TensorFlow 2.0, it uses the all-reduce algorithm to keep the neural network variables in sync during training.


\subsection*{Models Training Results}

\begin{figure}
\centering
\includegraphics[width=3in]{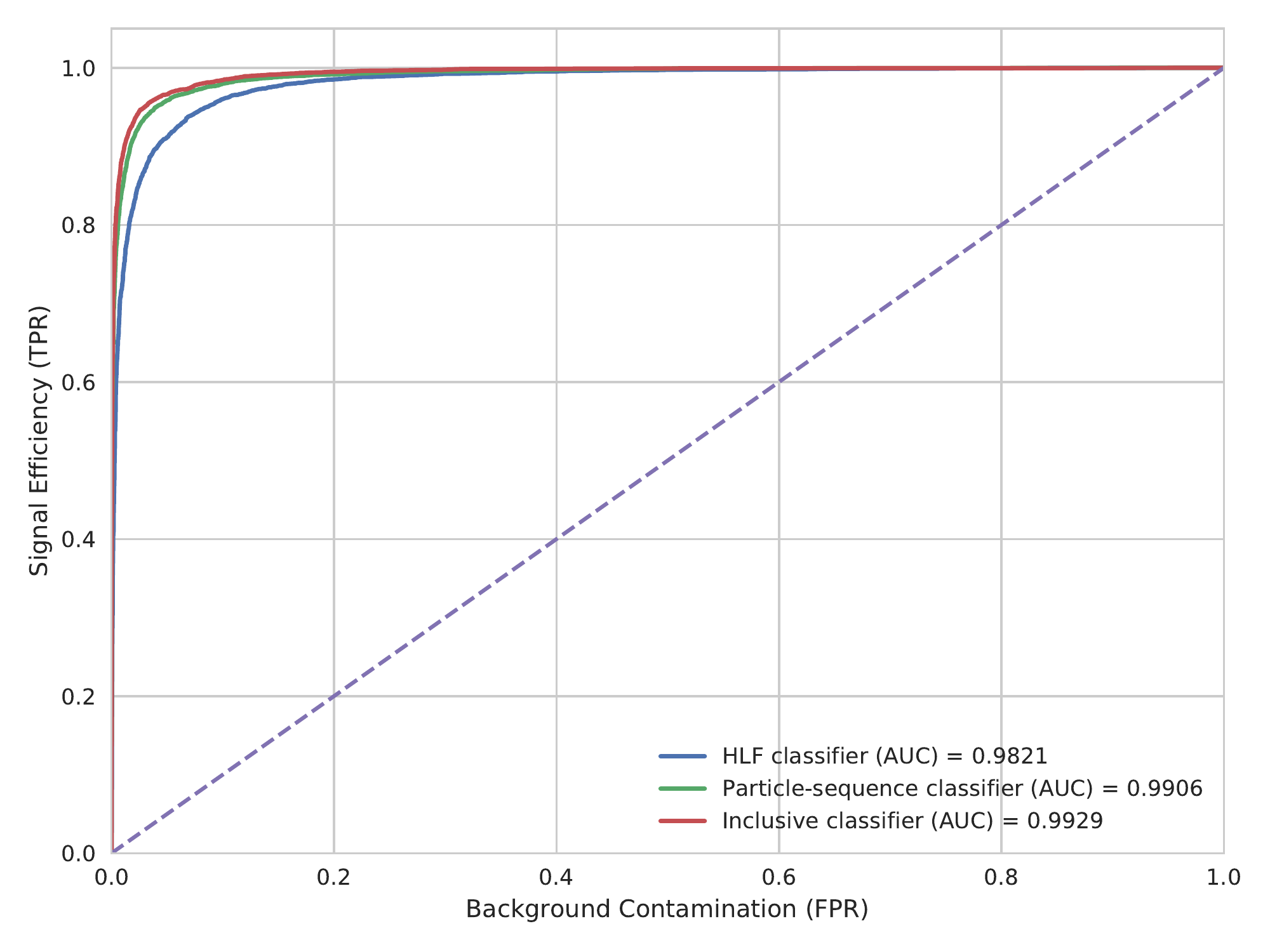}
\caption{AUC and ROC curves for the $t\bar{t}$ selector trained using BigDL. The results show that all three models perform well, with the inclusive classifier model being the best result of the three. This is matches the results of Nguyen et al. \cite{2018arXiv180700083N}}\label{ref:rocCurve}
\end{figure}

After training the three neural network models, we evaluated the results using the test dataset. Each of the models presented in the previous section returns as output the probability that an input event is associated with a given topology: $y_{QCD}$, $y_{W}$ or $y_{t\bar{t}}$. This can be used to define a classifier, for example, it suffice to apply a threshold requirement on $y_{t\bar{t}}$ or $y_{W}$ to define a $W$ or a $t\bar{t}$ classifier, respectively.
A common technique to evaluate the performance of classifiers, also utilized in the reference work \cite{2018arXiv180700083N}, is to compute and compare their corresponding ROC (receiver operating characteristic curve) curves and AUC (area under the ROC curve). \figurename{ \ref{ref:rocCurve}} shows the comparison of the ROC curves for the three classifiers for a $t\bar{t}$ selector.\\

\begin{figure}
\centering
\includegraphics[width=3in]{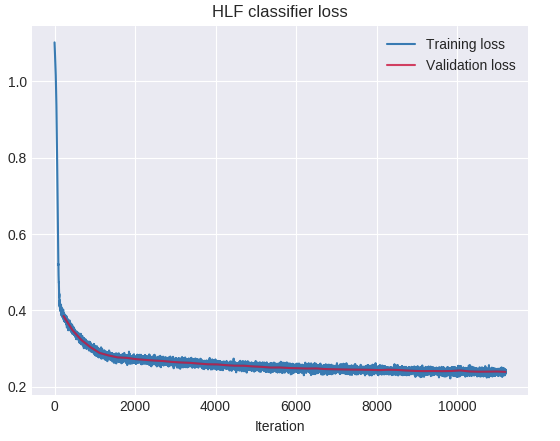}
\caption{Training and validation loss (categorical cross entropy) plotted as a function of iteration number for HLF classifier. Training convergence for the neural networks used in this work is obtained at around 50 epochs, depending on the model.}\label{ref:LossHLF}
\end{figure}

\noindent The fact that the HLF classifier performs well, despite its simplicity, can be justified by the fact that considerable physics knowledge is built into the definition of the 14 high-level features. This conclusion is further reinforced by the results of additional tests, where we have built the topology classifier using a \say{random forest} and a \say{gradient boosting} (XGBoost) model, respectively, and trained them using the high level features dataset. In both cases, we have obtained very good performance of the classifiers, just close to what has been achieved with the HLF classifier model based on feed forward neural networks.
In contrast, the fact that the particle sequence classifier performs better than the HLF classifier is remarkable, because we are not putting any {\it a priori} knowledge into the particle sequence classifier: the model just uses a list of reconstructed particles as input for the training data. In some sense, the GRU layer is identifying important features and physics quantities out of the training data, rather than by using knowledge injected via feature engineering. Further improvements are seen by combining the HLF classifier and the particle sequence classifier, we call the resulting model the "inclusive classifier" model.\\
\noindent Another remarkable fact is that the distributed training procedure used in this work converges to the same results presented in the original paper paper\cite{2018arXiv180700083N}. \figurename{ \ref{ref:LossHLF}} illustrates the smooth training convergence for the HLF classifier.

\subsection{Workload and Performance}

\subsubsection{Platforms and Hardware}
Multiple Spark clusters have been used to develop, run and test the data pipeline. The first group consisted of two Hadoop YARN clusters, part the CERN Hadoop and Spark service: one development cluster, and one production cluster consisting of 52 nodes, with the following resources: 1800 vcores, 14 TB of RAM, 9 PB of storage. The production cluster is a shared general-purpose multi-tenant Hadoop YARN cluster built using commodity hardware and running Linux (CentOS 7). Only a fraction of the resources of the production cluster capacity was used when executing the pipeline (up to about 30\% of the total core capacity). Jobs from other and different workloads were concurrently running on the system used to develop this work, which has introduced some noise and possibly impacted performance, however, this has also provided with a ``real-life'' scenario of what physicists could achieve when running their data pipelines and DL training jobs on a shared Spark cluster.  
Further work has been done using cloud resources, on-premises and at a public cloud. In particular, using Kubernetes clusters on the CERN Cloud Service and on Oracle Cloud Infrastructure have been tested. Analogously to the case of YARN tests, also in this case resources were allocated on a multi-tenant system, such as cloud systems, although with a notable difference that the allocated VM resources were not further shared, but used exclusively for this workload. When running Spark workloads on cloud resources, we have used Kubernetes as a cluster solution for Spark.

\subsubsection{Data Preparation}
Data ingestion and event filtering is a very resource-demanding step of the pipeline. Processing the original data set of 4.5 TB took approximately 3 hours when deployed on a Hadoop YARN cluster with 50 executors, each executor allocating 8 cores, for a total of 400 cores allocated to the Spark application.
The data ingestion and feature preparation workloads were monitored and measured using metrics from OS tools and the Spark metrics system. Monitoring data showed that the majority of the application time was spent by tasks running ``on CPU''. CPU cycles were spent mostly executing Python code, therefore outside of the Spark JVM. This can be explained by the fact that we chose to process the bulk of the training data using Python UDF functions. This a well-known behavior in current versions of Apache Spark when using Python UDF extensively. In such systems many CPU cycles are ``wasted'' in data serialization and deserialization operations, going back and forth from JVM and Python, in the current Spark implementation. 
Spark runs more efficiently when using the DataFrame API and/or Spark SQL. To improve the performance of the event filtering we have refactored part of the Python UDF code in the data ingestion step, by using Spark SQL functions. Notably, we have made use of Spark SQL Higher Order Functions, a specialized category of SQL functions, introduced in Spark from version 2.4.0, that allow improving processing for nested data (arrays) using SQL and Spark's Dataframe API. In the case of our workload, this has introduced the benefit of running a significant fraction of the filtering operations fully inside the JVM, optimized by the Spark code for DataFrame operations. The result is that the data ingestion step, optimized with Spark SQL and higher order functions, ran in about 2 hours, improving on previously measured job duration of 3 hours, obtained with the implementation that uses only Python UDF.
Future work on the pipeline may further address the issue of reducing the time spent in serialization and deserialization when using Python UDF, for example, we might decide to re-write the critical parts of the ingestion code in Scala. However, with the current training and test data size (4.5 TB of raw data), the performance of the data preparation steps, currently of the order of 3 hours for the combined data ingestion and feature preparation steps, is acceptable. 
Notably, running data preparation on the full dataset is in practice not the most time-consuming part of the process: development and testing on subsets of the full dataset, being typically where the physicists would spend most of their time for the data preparation steps. A standard and easy-to-use API to process data as scale, such as the one provided by Spark, can play an important role in helping the physicists to be more productive and ultimately improve the performance of the data preparation step.

\begin{figure}[t]
\centering
\includegraphics[width=3in]{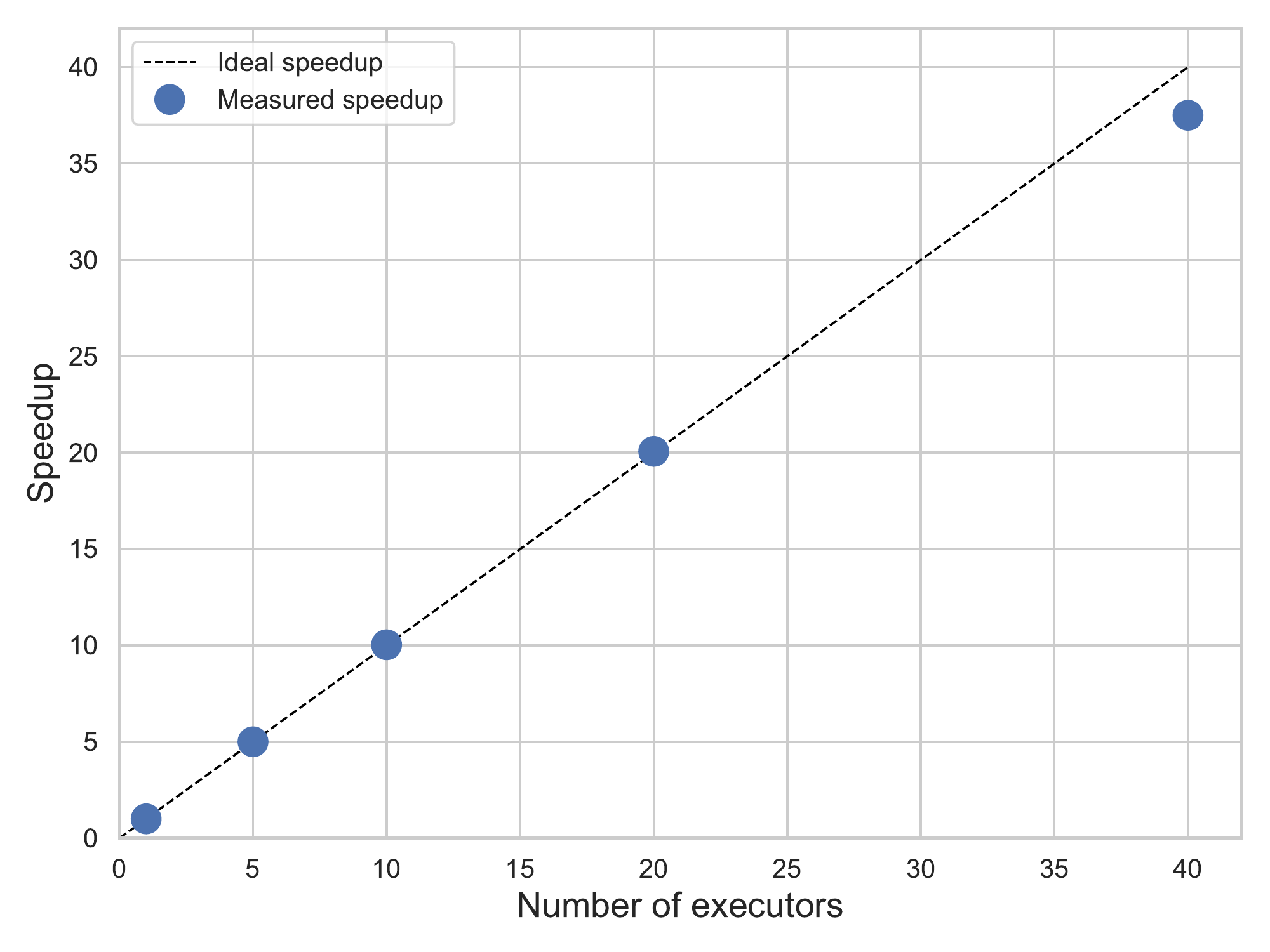}
\caption{Speedup of the grid search time for the High Level Features classifiers on $100\,\text{K}$ events and on a grid of $\sim 200$ parameters with 8-fold cross validation.}\label{ref:gridSearchTime}
\end{figure}

\subsubsection{Hyperparameter Tuning}
Performance of hyperparameter tuning with grid search: grid search runs multiple training jobs in parallel, therefore it is expected to scale well when run in parallel, as it was the case in our work (see also the paragraph on hyperparameter tuning). This was confirmed by measurements, as shown in \figurename{ \ref{ref:gridSearchTime}}, by adding executors, and consequently the number of models trained in parallel, the time required to scan the parameters space decreased, as expected.

\begin{figure}
\centering
\includegraphics[width=3in]{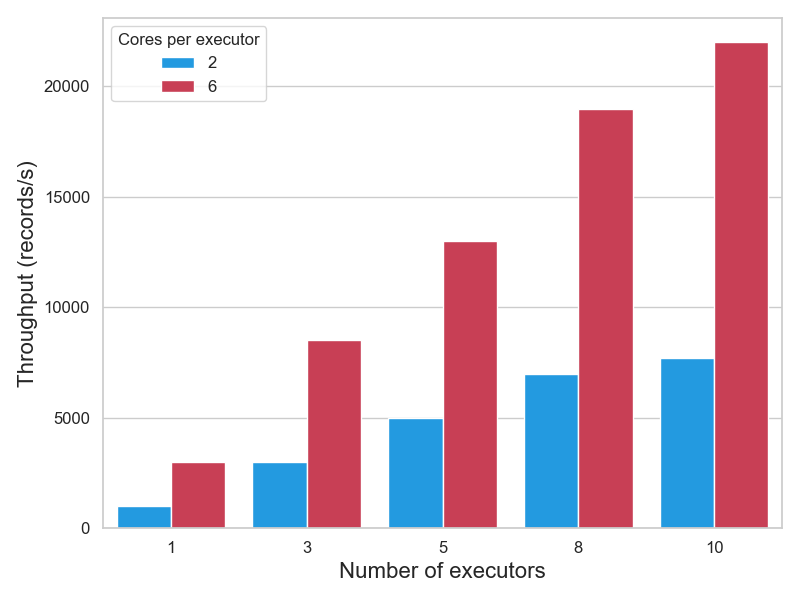}
\caption{Throughput of BigDL changing the number of executors and cores per executor for the HLF classifier.}
\label{ref:scalabilityBDL}
\end{figure}

\subsubsection{Training with BigDL}
Performance of distributed training with BigDL and Analytics Zoo is also important for this exercise, as faster training time means improved productivity of the physicists who will typically perform many experiments and fine-tuning on the models' structure and training parameters. \figurename{ \ref{ref:scalabilityBDL}} shows a few measurements of the training speed for the HLF classifier. The tests have been run on the development cluster, using batch size of 32 per worker, and show very good scalability behavior of the training at the scale tested. Additional measurements, using 20 executors, with 6 cores each (for a total of 120 allocated cores), using batch size of 128 per worker, showed training speed for the HLF classifier of the order of 100K rows/sec, sustained for the training of the full dataset. We were able to train the model in less than 5 minutes, running for 12 epochs (each epoch processing 3.4 million training events). We found that in our setup the batch size we used had an impact on the execution time and the accuracy of the training. We found that a batch size of 128 for the HLF classifier was a good compromise for speed, while a batch size of 32 was slower but gave improved results. We would use the former for model development and the latter, for example for producing the final training results. 

An important point to keep in mind when training models with BigDL, is that the RDDs containing the features and labels datasets need to be cached in the executors' JVM memory. This can be a significant amount of memory, of about 250 GB when training the "particle sequence" and "inclusive classifier" models, as they feed on training data with features containing large particle matrices (801 x 19).

\begin{figure}
\centering
\includegraphics[width=3in]{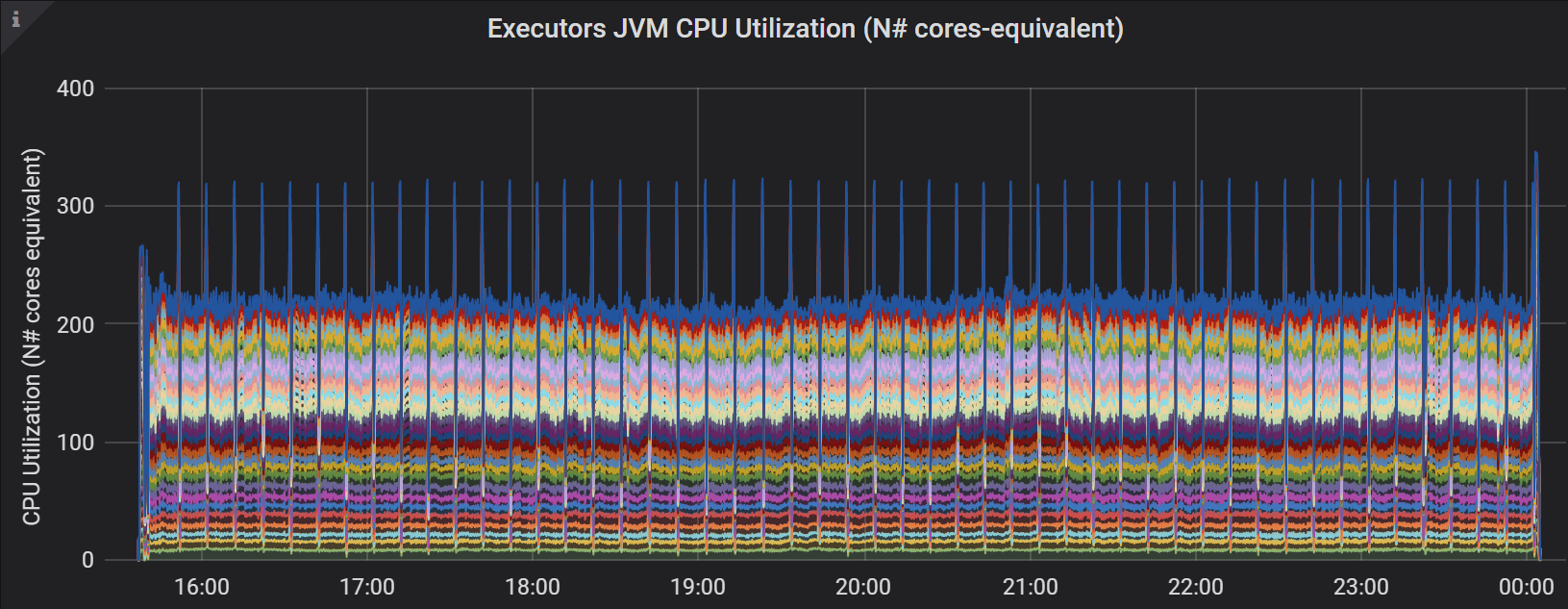}
\caption{Measured aggregated CPU utilization of Spark executors during the training of the Inclusive classifier. The neural network training has lasted for 8.5 hours and has utilized about 215 concurrently running CPU cores for the whole duration of the training job.}\label{ref:CPUMeasurement}
\end{figure}

The training dataset size and the model complexity of the HLF classifier are of relatively small scale, this makes the HLF classifier suitable for training on desktop computers, i.e. without using distributed computing solutions. In contrast, the Particle Sequence Classifier and the Inclusive Classifier models have much higher complexity and require processing hundreds of GB of training data. \figurename{ \ref{ref:CPUMeasurement}} shows the amount of CPU time consumed during the training of the Inclusive Classifier model using BigDL/Analytics-zoo on a Spark cluster, running on 30 executor instances (12 cores allocated for each executor) and training with a batch size of 32. Notably, neural network training, in this case, has lasted for 8.5 hours (50 epochs) and has utilized $6.6\ 10^6$ CPU seconds, this is means that, on average, 215 CPU cores were concurrently active for the duration of the training job. The executor CPU time measurements have been collected using Spark metrics instrumentation feeding into an InfluxDB instance, the monitoring plots have been generated using a Grafana dashboard, custom-built to visualize Spark monitoring data.
By measuring the training performance multiple times across several days and months, we noticed that the training performance in our setup depends on the state of the cluster and is affected by stragglers and executors under-performing because of hardware faults or high server load. 

\subsubsection{Training with TensorFlow}
We have also run tests and measured the distributed training performance of the Inclusive Classifier model using TensorFlow. In particular, we have used tf.keras and the module tf.distribute implementing ``multi worker mirror strategy'', to run and distribute the training over multiple workers. We have tested different configurations, varying the number of workers, servers, CPU cores, and GPU devices. 
Our tests with TensorFlow (tf.keras) used training and test data in TFRecord format, produced at the end of the data preparation part of the pipeline. TensorFlow reads natively TFRecord format and has tunable parameters and optimizations when ingesting this type of data using the modules tf.data and tf.io. In particular, we followed TensorFlow's documentation recommendations for improving the data pipeline performance, by using prefetching, parallel data extraction, sequential interleaving, and by using a large read buffer, and caching (in particular caching was used for distributed training when running on multiple nodes, to take advantage of the aggregated memory available in the cluster).

The first TensorFlow training tests we performed were deployed on a single bare metal machine, with 24 physical cores (Broadwell) and 512 GB of RAM. Training and test data were stored locally on an SSD, therefore minimizing time spent doing I/O. Training for 5 epoch, with batch size 128, using TensorFlow 2.0-rc0 using this system took 21 hours and reached results for loss and AUC are similar to what had been obtained in previous runs with BigDL (however, from further tests with this type of configuration using distributed training discussed later, we found that 12 epochs of training would provide the highest quality of training results).
While running the training experiments with TensorFlow on the bare metal machine, we noticed that the TensorFlow process performing the training would not utilize more than about 5 concurrent cores, despite the availability of idle CPU cores in the machine. External causes of this behavior, such as an I/O bottlenecks, could be ruled out; we explain the observed behavior as due to the way TensorFlow internally parallelizes its operations for training the Inclusive Classifier neural network on CPU, as this would naturally lead to a maximum number of concurrent threads executions, depending on the algorithms used, and may change in future versions.
To take advantage of the full 24 physical cores of the test machine, we, therefore, implemented distributed training, by using the tf.distribute module implementing the multi worker mirror strategy, and by manually starting multiple workers. We found that we could scale up to 4 concurrent training workers on the bare metal machine used for testing. Training of the Inclusive Classifier with 4 concurrent workers for 6 epochs, with batch size 128 (per worker) using TensorFlow 2.0.1 took 7.7 hours. 
We have also run training using TensorFlow on a dedicated test machine equipped with GPU. The first finding we found was striking differences in performance depending on the TensorFlow version. TensorFlow 2.0 has introduced an optimization for training GRU layers on GPU, which improves the performance of about 2 orders of magnitude compared to TensorFlow 1.14. We used a bare metal machine with NVidia P100, running TensorFlow 2.0-rc1 with CUDA 10.0, with the training and test data stored locally on the machine filesystem. Training the Inclusive Classifier took 2.3 hours for 6 epochs with batch size 128, producing very good results for the trained classifier. We have also trained using the same hardware and software setup, i.e. using TensorFlow 2.0-rc0 and an NVidia P100 GPU, a neural network for the Inclusive Classifier with a small modification, consisting in replacing the GRU layer with an LSTM layer. The neural network training was run for 6 epochs, using batch size 128. The training time in this case was 1.5 hours, with the achieved loss and AUC comparable to the results obtained with the GRU model.
Previously, a model based on LSTM was discarded by the authors of \cite{2018arXiv180700083N}, and the GRU layer was chosen for performance reasons. We find that conclusion to be still applicable in our training tests on CPU resources. We believe that the performance improvement we measured using LSTM on GPU training tests, is due to an optimization introduced in the implemen-tation of LSTM layers for CUDA/GPU in the version of TensorFlow used in our tests.

\subsubsection{TensorFlow on Kubernetes}

Cloud resources provide a suitable environment for scaling distributed training of neural networks. One of the key advantages of using cloud resource is the elasticity of the platform that allows allocating resources when needed. Moreover, container orchestration systems, in particular Kubernetes, provide a powerful and flexible API for deploying many types of workloads on cloud resources, including machine learning and data processing tools. CERN physicists can access cloud resources and Kubernetes clusters via the CERN private cloud. The use of public clouds is also being actively tested for HEP workloads. In particular, the tests reported here have been run using resources from Oracle's OCI. 
For this work, we have developed a custom launcher script (TF-Spawner \cite{tfspawner}) for running distributed TensorFlow training code on Kubernetes clusters. TF-Spawner takes as input the Python code used for TensorFlow training described earlier, which makes use of tf.distribute for distributing the training, of tf.data for the data pipeline, and it runs the code on a Kubernetes cluster using container images. We used the official TensorFlow images from Docker Hub for these tests. Moreover, TF-Spawner takes care of distributing the necessary credentials for authenticating with cloud storage and environment variables needed by tf.distribute, it takes care of allocating the desired number of workers, as pods (units of execution) on a Kubernetes cluster, and of cleaning up the resources.
Training and test data had been copied to the cloud object storage prior to running the tests, notably copying the TFRecords files to the OCI object storage (for tests run at CERN we used the S3 object storage). Reading from OCI storage can become a bottleneck for distributed training, as it requires reading data over the network which can suffer from bandwidth saturation, latency spikes and/or multi-tenancy noise. TensorFlow's optimizations for the data pipeline performance discussed earlier were applied to these tests too. Notably, caching has proven to be very useful for distributed training with GPUs and for some of the largest tests on CPU, as we observed that in those cases the first training epoch, which has to read the data into the cache, was much slower than subsequent epoch which would find data already cached.
Tests discussed in this section were run using TensorFlow version 2.0.1, using tf.distribute strategy ``multi worker mirror strategy''. Additional care was taken to make sure that the different tests would also yield the same good results in terms of accuracy on the test dataset as what was found with previously tested training methods. To achieve this we have found that additional tuning was needed on the settings of the learning rate for the optimizer (we use the Adam optimizer for all the tests discussed in this article). We scaled the learning rate with the number of workers, to match the increase in effective batch size (we used 128 for each worker), this is a well-known technique and it is described for example in \cite{LargeBatchSGD}. In addition, we found that slowly reducing the learning rate as the number of epochs progressed, was beneficial to the convergence of the network. This additional step is an ad hoc tuning that we developed by trial and error and that we validated by monitoring the accuracy and loss on the test set at the end of each training.
To gather performance data, we ran the training for 6 epochs, which provided accuracy and loss very close to the best results reported earlier, while optimizing on the time needed to take the measurements. Similarly to what we observed with previous tests, we also confirmed for this case that training the network up to 12 epochs provides better results for accuracy and loss.
We have also tested adding shuffling between each epoch, using the shuffle method of the tf.data API, however this has not shown measurable improvements so it has not been further used in the tests reported here.

\begin{figure}[h]
\centering
 \includegraphics[width=3in]{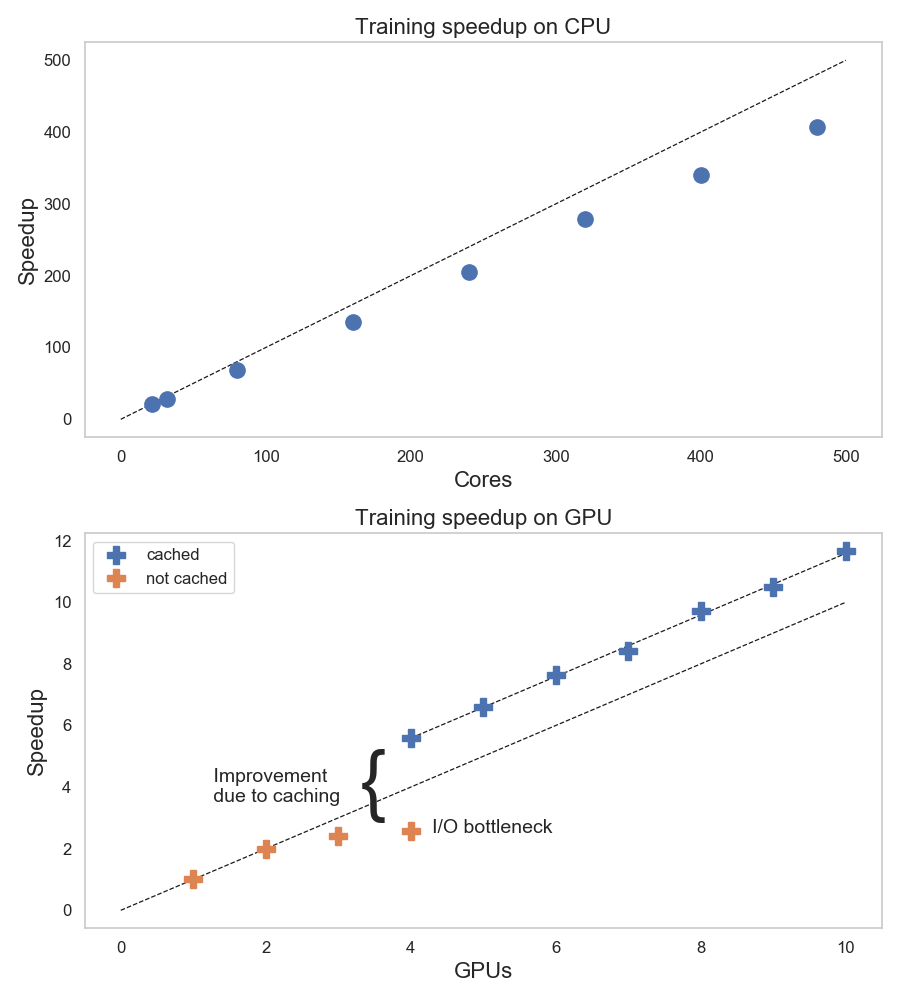}
\caption{Measured speedup for the distributed training of the Inclusive Classifier model using TensorFlow and tf.distribute with ``multi worker mirror strategy'', running on cloud resources with CPU and GPU nodes, training for 6 epochs. The speedup values indicate how well the distributed training scales as the number of worker nodes with CPU and/or GPU resources increases.}
\label{ref:TensorFLowDistributedSpeedup}
\end{figure}

\figurename{ \ref{ref:TensorFLowDistributedSpeedup}} shows the results of the Inclusive Classifier model training speedup for a variable number of nodes and CPU cores. Measurements show that the training time decreases as the number of allocated cores is increased. The speedup grows close to linearly in the range tested: from 32 to 480 cores.
Tests used resources from Oracle's OCI, where we built a Kubernetes cluster using virtual machines (VMs) and configured it with a set of Terraform script to automate the process. The cluster for CPU tests used VMs of the flavor ``VM.Standard2.16'', based on 2.0 GHz Intel Xeon Platinum 8167M, each providing 16 physical cores (Oracle cloud refers to this as OCPUs) and 240 GB of RAM. Tests in this configuration deployed 3 pods for each VM, each pod running one TensorFlow worker taking part in the distributed training cluster. Additional OS-based measurements on the VMs confirmed that this was a suitable configuration, as we could measure that the CPU utilization on each VM matched the number of available physical cores (OCPUs), therefore providing good utilization without saturation. The available RAM in the worker nodes was used to cache the training dataset using the tf.data API (data populates the cache during the first epoch).

\begin{figure}[h]
\centering
 \includegraphics[width=3in]{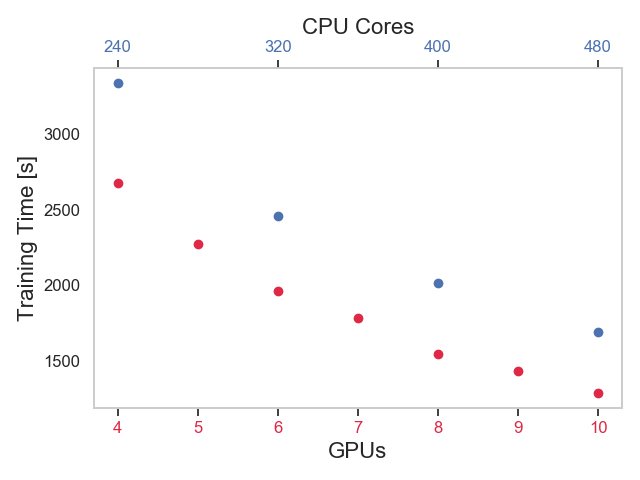}
\caption{Selected measurements of the distributed training time for the Inclusive Classifier model using TensorFlow and tf.distribute with ``multi worker mirror strategy'', training for 6 epochs, running on cloud resources using CPU and GPU nodes.}
\label{ref:TensorFLowDistributedTrainingTime}
\end{figure}

Similarly, we have performed tests using GPU resources on OCI and TF-Spawner. For the GPU tests we have used the VM flavor ``GPU 2.1'' which comes equipped with an Nvidia P100 GPU, 12 physical cores (OCPU) and 72 GB of RAM. We have tested distributed training with up to 10 GPUs and found almost linear scalability in the tested range. One important lesson learned is that when using GPUs the slow performance of reading data from OCI storage makes the first training epoch much slower than the rest of the epochs (up to 3-4 times slower). It was therefore very important to use TensorFlow's caching for the training dataset for our tests with GPUs, however, we could only do that for tests with 4 nodes or more, given the limited amount of memory in the VM flavor used (72 GB of RAM per node) compared to the size of the training set (200 GB).
Distributed training tests with CPUs and GPUs were performed using the same infrastructure, namely a Kubernetes cluster built on cloud resources and cloud storage allocated on OCI, moreover, we used the same script for training with tf.distribute and tf.keras, and the same TensorFlow version. \figurename{ \ref{ref:TensorFLowDistributedTrainingTime}} shows the distributed training time measured for some selected cluster configurations. We can use these results to compare the performance we found when training on GPU and on CPU. For example, we find there that the time to train the Inclusive Classifier for 6 epochs using 400 CPU cores (distributed over 25 VMs equipped with 16 physical cores each) is about 2000 seconds, which is similar to the training time we measured when distributing the training over 6 nodes equipped with GPUs. We do not believe these results can be easily generalized to other environments and models, however, they are reported here as they can be useful as an example and future reference.
When training using GPU resources (Nvidia P100), we measured that each batch is processed in about 59 ms (except for epoch 1 which is I/O bound and is about 3x slower). Each batch contains 128 records, and has a size of about 7.4 MB. This corresponds to a measured throughput of training data flowing through the GPU of about 125 MB/sec per node (i.e. 1.2 GB/sec when training using 10 GPUs). When training on CPU, the measured processing time per batch is about 930 ms, which corresponds to 8 MB/sec per node, and amounts to 716 MB/sec for the training test with 90 workers and 480 CPU cores.



\section*{Conclusions and Future Outlook}

This work shows an example of a pipeline for end-to-end data preparation and deep learning for a high energy physics use case and details of how it can be implemented using tools and techniques from open source projects and ``big data'' communities at large.
In particular, it addresses the implementation of a pipeline for data preparation and training of a particle topology classifier based on deep learning. Following the work and models developed by Nguyen et al. \cite{2018arXiv180700083N},
event topology classification, based on deep learning, is used to improve the purity of data samples selected at the trigger level.
The application of the classifier developed in this work is intended at improving the efficiency of LHC detectors and data flow systems for data acquisition. The application of the methods and techniques developed in this paper, using open source and big data tools,
is intended to improve scientist productivity and resource utilization when working on data analysis and machine learning model development.

Machine learning and deep learning on large amounts of data are standard tools for particle physics, and their use is expected to increase in the HEP community in the coming year, both for data acquisition and for data analysis workflows, notably in the context of the challenges of the High Luminosity LHC project\cite{HighLumi}.
Improvements in productivity and cost reduction for development, deployment and maintenance of machine learning pipelines on HEP data are of high interest.
The authors believe that using tools and methods from  open source and big data communities at large, brings important advantages in this area, in particular in terms 
of usability and developers' productivity. Key advantages are the use of standard APIs supported by a large community,
and ease of deployment and integration with modern computing systems, notably cloud systems.

The results discussed in this paper provide insights on how to perform data preparation at scale and how to use scale-out cluster resources like YARN and Kubernetes. Moreover, we showed how one can easily deploy distributed training on cloud resources both using CPUs and GPUs.  We expect that many of the tools and methods discussed here will evolve considerably in the near future, following the directions
taken by the relevant communities. This will most likely make the details of the implementations discussed in this paper obsolete, 
however, we can also expect that the evolution will bring important improvements and will profit greatly from
the experience and lessons learned by a large user and developer base.

Reference notebooks with code developed for this work and data samples are available at: https://github.com/cerndb/SparkDLTrigger

The authors would like to thank the authors of \cite{2018arXiv180700083N} and in particular Thong Nguyen and Maurizio Pierini for their help and suggestions,
the team of CERN Spark, Hadoop and streaming service, CERN openlab, in particular Maria Girone, Viktor Khristenko, Michal Bien, the team of CMS Bigdata project, in particular Oliver Gutsche, Jim Pivarski, Matteo Cremonesi and the BigDL and Analytics Zoo team at Intel, in particular Jiao (Jennie) Wang and Sajan Govindan.

\bibliographystyle{ieeetr}
\bibliography{Bibliography}{}

\begin{thebibliography}{10}

\bibitem{Zaharia:2010:SCC:1863103.1863113}
M.~Zaharia, M.~Chowdhury, M.~J. Franklin, S.~Shenker, and I.~Stoica, ``Spark:
  Cluster computing with working sets,'' in {\em Proceedings of the 2Nd USENIX
  Conference on Hot Topics in Cloud Computing}, HotCloud'10, (Berkeley, CA,
  USA), pp.~10--10, USENIX Association, 2010.

\bibitem{Baldi:2014kfa}
P.~Baldi, P.~Sadowski, and D.~Whiteson, ``{Searching for Exotic Particles in
  High-Energy Physics with Deep Learning},'' {\em Nature Commun.}, vol.~5,
  p.~4308, 2014.

\bibitem{2018arXiv180405839D}
J.~{Dai}, Y.~{Wang}, X.~{Qiu}, D.~{Ding}, Y.~{Zhang}, Y.~{Wang}, X.~{Jia},
  C.~{Zhang}, Y.~{Wan}, Z.~{Li}, J.~{Wang}, S.~{Huang}, Z.~{Wu}, Y.~{Wang},
  Y.~{Yang}, B.~{She}, D.~{Shi}, Q.~{Lu}, K.~{Huang}, and G.~{Song}, ``{BigDL:
  A Distributed Deep Learning Framework for Big Data},'' {\em arXiv e-prints},
  p.~arXiv:1804.05839, Apr 2018.

\bibitem{2018arXiv180700083N}
T.~Q. {Nguyen}, I.~{Weitekamp}, Daniel, D.~{Anderson}, R.~{Castello},
  O.~{Cerri}, M.~{Pierini}, M.~{Spiropulu}, and J.-R. {Vlimant}, ``{Topology
  classification with deep learning to improve real-time event selection at the
  LHC},'' {\em Comput Softw Big Sci (2019) 3: 12}, August 2019.

\bibitem{BRUN199781}
R.~Brun and F.~Rademakers, ``Root — an object oriented data analysis
  framework,'' {\em Nuclear Instruments and Methods in Physics Research Section
  A: Accelerators, Spectrometers, Detectors and Associated Equipment},
  vol.~389, no.~1, pp.~81 -- 86, 1997.
\newblock New Computing Techniques in Physics Research V.

\bibitem{Bird:1695401}
I.~Bird, P.~Buncic, F.~Carminati, M.~Cattaneo, P.~Clarke, I.~Fisk, M.~Girone,
  J.~Harvey, B.~Kersevan, P.~Mato, R.~Mount, and B.~Panzer-Steindel, ``{Update
  of the Computing Models of the WLCG and the LHC Experiments},'' Tech. Rep.
  CERN-LHCC-2014-014. LCG-TDR-002, Apr 2014.

\bibitem{TMVA}
A.~{Hoecker}, P.~{Speckmayer}, J.~{Stelzer}, J.~{Therhaag}, E.~{von Toerne},
  H.~{Voss}, M.~{Backes}, T.~{Carli}, O.~{Cohen}, A.~{Christov}, D.~{Dannheim},
  K.~{Danielowski}, S.~{Henrot-Versille}, M.~{Jachowski}, K.~{Kraszewski},
  J.~{Krasznahorkay}, A., M.~{Kruk}, Y.~{Mahalalel}, R.~{Ospanov},
  X.~{Prudent}, A.~{Robert}, D.~{Schouten}, F.~{Tegenfeldt}, A.~{Voigt},
  K.~{Voss}, M.~{Wolter}, and A.~{Zemla}, ``{TMVA - Toolkit for Multivariate
  Data Analysis},'' {\em arXiv e-prints}, p.~physics/0703039, Mar 2007.

\bibitem{EOS}
A.~J~Peters and L.~Janyst, ``Exabyte scale storage at cern,'' {\em Journal of
  Physics: Conference Series}, vol.~331, 12 2011.

\bibitem{viktor_khristenko_jim_pivarski_2017}
V.~Khristenko and J.~Pivarski, ``diana-hep/spark-root: Release 0.1.14,'' Oct
  2017.

\bibitem{xRootDconnector}
CERN-DB, ``Hadoop-xrootd connector.'' https://github.com/cerndb/hadoop-xrootd,
  2013.

\bibitem{HadoopYARN}
``{Apache Hadoop Project}.'' \url{https://hadoop.apache.org/}.

\bibitem{ApacheParquet}
``{Apache Parquet}.'' \url{https://parquet.apache.org/}.

\bibitem{protobuf}
Google, ``Protocol buffers.''
  \url{http://code.google.com/apis/protocolbuffers/}.

\bibitem{scikitlearn}
``Scikit-learn.'' \url{https://scikit-learn.org/}.

\bibitem{kerastuner}
``Keras tuner.'' \url{https://keras-team.github.io/keras-tuner/}.

\bibitem{Kubernetes}
``Kubernetes.'' \url{https://kubernetes.io/}.

\bibitem{chollet2015keras}
F.~Chollet {\em et~al.}, ``Keras.'' https://keras.io, 2015.

\bibitem{openlab}
``{CERN} openlab.'' \url{https://openlab.cern/}.

\bibitem{AnalyticsZoo}
``Analytics zoo.'' \url{https://analytics-zoo.github.io/}.

\bibitem{Tensorflow}
``Tensorflow.'' \url{https://www.tensorflow.org/}.

\bibitem{tfspawner}
``{TF-Spawner}.'' \url{https://github.com/cerndb/tf-spawner}.

\bibitem{LargeBatchSGD}
P.~{Goyal}, P.~{Doll{\'a}r}, R.~{Girshick}, P.~{Noordhuis}, L.~{Wesolowski},
  A.~{Kyrola}, A.~{Tulloch}, Y.~{Jia}, and K.~{He}, ``{Accurate, Large
  Minibatch SGD: Training ImageNet in 1 Hour},'' {\em arXiv e-prints},
  p.~arXiv:1706.02677, Jun 2017.

\bibitem{HighLumi}
``{High Luminosity LHC Project}.'' \url{http://hilumilhc.web.cern.ch}.

\end{thebibliography}
 
\end{document}